%% file: Main.tex
\documentclass{vgtc}                     

\ifpdf
  \pdfoutput=1\relax                   
  \usepackage{graphicx}                
  \DeclareGraphicsExtensions{.pdf,.png,.jpg,.jpeg} 
\else
  \usepackage{graphicx}                
  \DeclareGraphicsExtensions{.eps}     
\fi%

\graphicspath{{figures/}{pictures/}{images/}{./}} 

\usepackage{microtype}                 
\PassOptionsToPackage{warn}{textcomp}  
\usepackage{textcomp}                  
\usepackage{mathptmx}                  
\usepackage{times}                     
\usepackage{cite}                      
\usepackage{tabu}                      
\usepackage{booktabs}                  
\usepackage{dblfloatfix}

\onlineid{1071}

\vgtccategory{empirical study}

\title{
Towards Autocomplete Strategies for Visualization Construction 
}

\author{%
  \authororcid{Wei Wei}{0000-0003-0780-1276},
   \authororcid{Samuel Huron}{0000-0002-9319-8559}, and 
  \authororcid{Yvonne Jansen}{0000-0001-5092-551X}
}
\author{Wei Wei \thanks{e-mail: weiwei@uvic.ca}\\ %
        \parbox{1.4in}{\scriptsize \centering University of Victoria \\ University of Calgary \\ Télécom Paris}
\and Samuel Huron\thanks{e-mail: samuel.huron@telecom-paris.fr}\\ %
     \scriptsize Télécom Paris, CNRS i3, Institut Polytechnique de Paris %
\and Yvonne Jansen\thanks{e-mail: yvonne.jansen@cnrs.fr}\\ %
     {\scriptsize Univ.~Bordeaux, CNRS, Inria, LaBRI}}
     
\teaser{
\centering
    \includegraphics[width=0.8\linewidth]{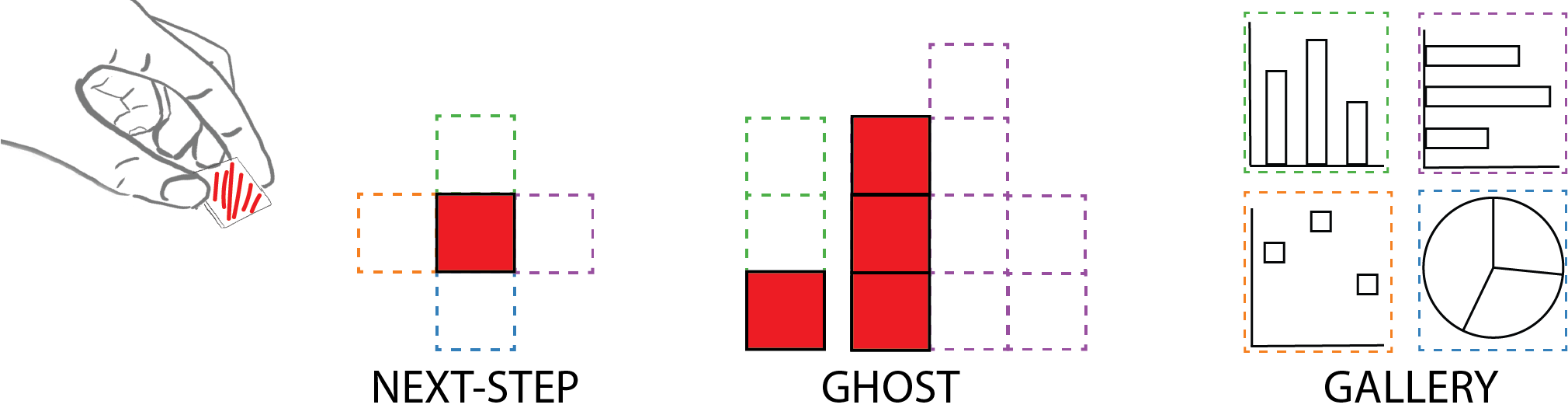}
\caption{ The three \textit{visualization autocomplete} strategies identified in our study. The dashed boxes in different colors represent autocomplete suggestions. Based on the token(s) that has been placed, \textsc{Next-Step} provides suggestions for a single next visual mapping operation. \textsc{Ghost} provides situated partial or completed visualization recommendations. \textsc{Gallery} provides a gallery of completed visual mapping options. }
\label{fig:teaser}
}

\abstract{
Constructive visualization uses physical data units - tokens -  to enable non-experts to create personalized visualizations engagingly. However, its physical nature limits efficiency and scalability. One potential solution to address this issue is autocomplete. By providing automated suggestions while still allowing for manual intervention, autocomplete can expedite visualization construction while maintaining expressivity. We conduct a speculative design study to examine how people would like to interact with a visualization authoring system that supports autocomplete. Our study identifies three types of autocomplete strategies and gains insights for designing future visualization authoring tools with autocomplete functionality.
  A free copy of this paper and all supplemental materials are available on our online repository: \url{https://osf.io/nu4z3/?view_only=594baee54d114a99ab381886fb32a126}.
}


\CCScatlist{
  \CCScatTwelve{Human-centered computing}{Visu\-al\-iza\-tion}{}{}
}

\begin{document}

\newcommand{\sam}[1]{\textcolor{blue}{[Sam] #1}}
\newcommand{\wei}[1]{\textcolor{brown}{[Wei] #1}}
\newcommand{\yvonne}[1]{\textcolor{green!70!black}{[Yvonne] #1}}
\newcommand{\todo}[1]{\textcolor{red}{#1}}

\newcommand{\nextstep}[1]{\textcolor{teal}{#1}}
\newcommand{\ghost}[1]{\textcolor{pink}{#1}}
\newcommand{\gallery}[1]{\textcolor{purple}{#1}}


\firstsection{Introduction}
\maketitle

Constructive visualization is a visualization authoring paradigm that emphasizes the act of manually assembling tokens mapped to data~\cite{huron2014constructive,huron2014constructing, wun2016comparing}. Its accessible method enables non-experts to build novel visualizations in an engaging way. 

While the process of manual assembly facilitates expressivity (the support for diverse visualization types), its ability to scale to thousands or millions of data points is limited. 
In the field of visualization, there has been a long-standing interest in using automation to assist in creating scalable visualizations in an efficient manner~\cite{Mackinlay1986Automating, MackinlayShowMe2007, viegas2007manyeyes, SatyanarayanReactiveVega2016}. 
In this paper, we are interested in how to apply automation mechanisms to constructive visualization while maintaining expressivity.
To achieve this, we explore autocomplete as a way to assist visualization construction that supports both automation and manual assembly. 
Autocomplete~\cite{bastTypeLessFind2006} is a widespread mechanism that aids in quickly obtaining desired outcomes based on user input. 
It supports manual operation while providing automated suggestions which can be accepted to speed up interaction or ignored to continue manually. Although originally used for text entry, autocomplete is now employed in various fields (~\cite{abiteboulAutocompletionLearningXML2012, khoussainovaSnipSuggestContextawareAutocompletion2010, Lo2019AutoFritz,bennettSimpleFlowEnhancingGestural2011}). 
Inspired by these works, 
we define \textit{visualization autocomplete} as \textbf{a function or strategy that supplies one or more options of predicted partial or complete visualization,} which could include any step of the visualization system of the infovis pipeline~\cite{jansen2013interaction}. We conducted a speculative design study to explore how autocomplete could be designed in the context of constructive visualization. 

One could imagine different ways to technically enable autocomplete mechanisms for constructive visualization (e.g., digital tokens, swarm bots~\cite{LeGocZooids2016} or XR technologies).
However, explicitly considering all these technologies constrains the potential interaction design and might introduce some level of indirection~\cite{Beaudouin-LafonDesigningInteraction2004}. Therefore, we focus here on exploring interaction design rather than the technology enabling potential interactions. Consequently, we asked 15 participants to sketch how they might want to interact with a \textit{visualization autocomplete} system without considering any technical constraints. 

Participants generated 93 pages of sketches which we analyzed through a thematic analysis. We identified three types of autocomplete strategies (shown in \autoref{fig:teaser}) that conclude how a system could provide visual mapping recommendations and automation of the visualization rendering during the construction process. We then illustrate each strategy in how they help people explore the visual mapping decision tree.
We further discuss the tradeoff between automation and expressivity embodied in the three strategies and potential schemes to facilitate better visualization construction.
Our work is a first attempt to explore how a system can support visual mapping autocomplete while enabling direct manipulation of the visual variables. We contribute by (1) proposing the first definition of \textit{visualization autocomplete}, (2) identifying three interaction design strategies, and (3) discussing future research and design possibilities.



\input{RelatedWork}
\input{Co-DesignStudy}

\input{Results}

\input{Discussion}
\input{Conclusion}

\acknowledgments{
The authors wish to thank Ehud Sharlin and Marcus Friedel. This work was supported by a grant from Mitacs and by the French government funding for the Future program (PIA) grant ANR-21-ESRE-0030 (CONTINUUM). }

\bibliographystyle{abbrv-doi}
\bibliography{Main}
\end{document}

%% file: RelatedWork.tex
\section{Related Work}
\label{Sect2}
\subsection{Autocomplete}
Autocomplete systems were initially designed to speed up text entry by providing suggestions for completing users' input in specialized input fields such as Unix command-line fields~\cite{korvemaker2000predicting}, file location fields~\cite{myersPresentFutureUser2000} and email address fields~\cite{myersPresentFutureUser2000}. Their applications then expanded to include word and phrase completion in general text editors such as Microsoft Word. Autocomplete also goes beyond text entry and is applied in various fields of information science to improve efficiency (e.g., programming assistance~\cite{abiteboulAutocompletionLearningXML2012,khoussainovaSnipSuggestContextawareAutocompletion2010},  breadboard design~\cite{Lo2019AutoFritz}, and enhancing gestural interaction~\cite{bennettSimpleFlowEnhancingGestural2011}). In general, autocomplete is ``\textit{a widely used mechanism to get to a desired piece of information quickly and with as little knowledge and effort as possible}~\cite{bastTypeLessFind2006}''. 

In the field of information visualization, a limited number of studies have explored the use of autocomplete for visual analysis as well as building visualization. Aiming to help the data discovery process, Setlur et al.~\cite{setlurSneakPiqueExploring2020} developed a system that incorporated text- and widget-based autocompletion to support query formulation in natural language interfaces for visual analysis. 
Another example, VisComplete~\cite{koop2008viscomplete}, provides assisted construction of visualization pipelines. When a user is adding a module to the pipeline, VisComplete suggests a completion. 

\subsection{Constructive Visualization}
There is a continuum between two ends that describe the approaches employed by visualization authoring tools~\cite{mendez2017bottom,huron2014constructing}: \textbf{top-down} and \textbf{bottom-up}. When building a visualization, top-down approaches start by defining high-level abstractions (e.g., visualization types, dimensions, and data mapping) and then populate them with data. Examples include the `Recommended Charts' in Microsoft Excel and the visualization browser in Voyager~\cite{wongsuphasawat2015voyager}. In contrast, bottom-up approaches start with individual data points and then build progressively towards defining higher-level structures, such as axes. A vital concept related to bottom-up approaches is constructive visualization - ``\textit{the act of constructing a visualization by assembling blocks, that have previously been assigned a data unit through a mapping}''~\cite{huron2014constructive}. 

One digital implementation that employs the paradigm of constructive visualization is iVoLVER~\cite{mendez2016ivolver}.
a web-based constructive visualization authoring tool that supports the manipulation of atomic elements of data. 
The authors of iVoLVER conducted a 
comparison study between iVoLVER (with bottom-up approaches) and Tableau (with top-down approaches)~\cite{mendez2017bottom}, and an investigation of how to reconcile automation with the benefits of constructive visualization~\cite{mendez2018considering}.

Our work also aims to facilitate constructive visualization through automation while maintaining expressivity. However, instead of implementing and studying a specific system, we study how people might want to interact with an ideal system. This approach is inspired by studying the interaction, not the interface~\cite{Beaudouin-LafonDesigningInteraction2004}. To focus on interaction at the level of the visual mark~\cite{saket2016visualization,saket2019investigating}, we used similar tangible tokens as previous constructive visualization studies and activities~\cite{huron2014constructing,wun2016comparing,willett2016constructive}. Furthermore, the simplicity of tokens allows for early exploration of various design alternatives, making them a powerful tool for our preliminary exploration~\cite{TohidiRightDesign2006}, and it is probably one of the earliest information processing tools human have used~\cite{schmandt2010writing}.



%% file: Co-DesignStudy.tex
\section{The Study}
We aim to understand how people might want to interact with a system able to autocomplete their visualization. To study the desired interactions without influencing our participants towards specific interfaces or technology, we decided to run a study in the form of a speculative design workshop~\cite{auger2013speculative}. The study consisted of four design workshops. Each workshop was run with a group of 3-5 participants.

\subsection{Participants}
We recruited 15 participants from a local university (8 females and 7 males. 7 aged 18-24 and 8 aged 25-44).  
We requested participants to complete a pre-study demographic survey, where we gathered information regarding their proficiency and frequency of metrics associated with visualization, interaction design, and XR technology. We inquired about XR technology to investigate the potential impact of participants' XR experience on their sketches. The demographic data are available in our online repository. Each workshop consisted of 3 sessions: \textbf{Construction}, \textbf{Ideation}, and \textbf{Discussion}. The participants carried out the first two alone and were involved in a group discussion in the last one. Each workshop took about 1.5 hours.

\begin{figure}[htbp]
\centering
\includegraphics[width = \linewidth]{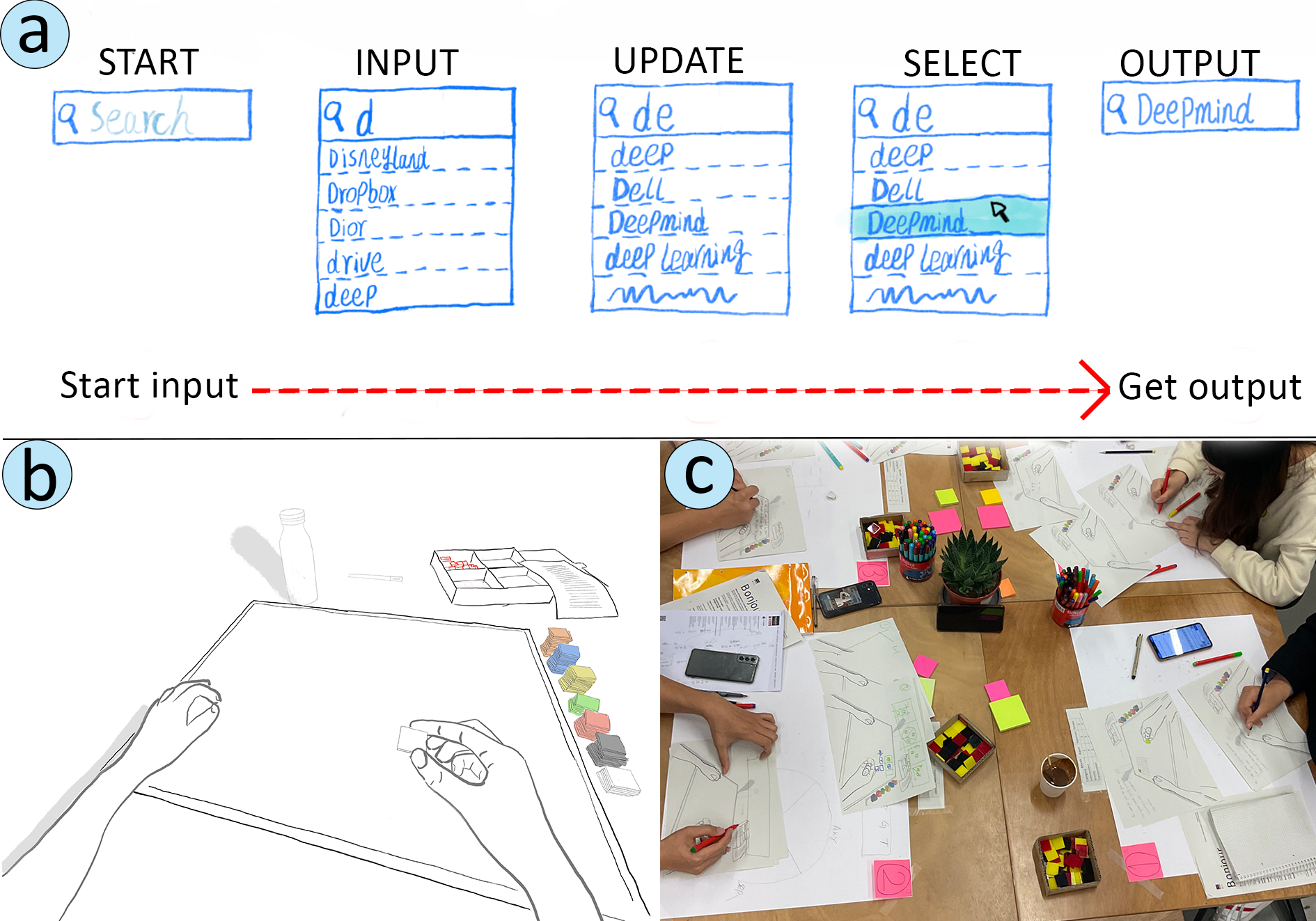}
\caption{(a) The diagram illustrates the mechanism of the search engine autocomplete feature. (b) The sketch template. (c) Participants were sketching their ideas.}
\label{Sect3-Figure2}
\end{figure}

\begin{figure*}[hb]
\centering
\includegraphics[width = \linewidth]{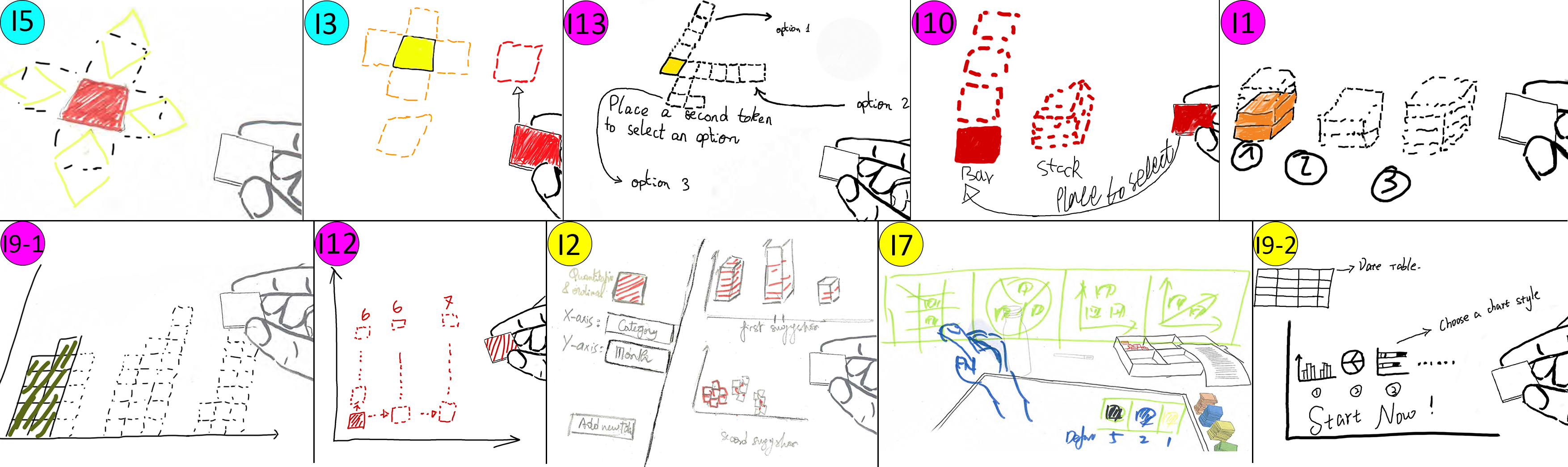}
\caption{The participants' design sketches illustrate three autocomplete strategies. I5 and I3 showcase the \textsc{Next-Step} strategy, highlighting available surrounding positions for the next token. The \textsc{Ghost} strategy is more flexible, providing suggestions for 2D (I13, I9-1, I12) and 3D (I10, I1) visualizations, both partial (I13, I10) and full (I1, I9-1, I12). And its suggestions can be based on one token (I13, I10, I12) or multiple tokens (I1, I9-1). Additionally, I2, I7, and I9-2 demonstrate the \textsc{Gallery} strategy, featuring multiple displays after reading data (I9-2) or defining high-level abstractions (I2, I7).}.
\label{Sect4-Figure1}
\end{figure*}
\label{Sect3}

\subsubsection{Construction}
The construction session introduced participants to the concept of constructive visualization. The setup followed Huron et al.~\cite{huron2014constructing}'s study (including a box of tokens, a printed dataset, and an A2 paper canvas). We simplified the data to positive integers up to eight to eliminate the need for explicit token mapping, which gave participants more time for manual assembling instead of mathematical calculations. At the workshop's outset, participants were informed of a person seeking their advice on expenses. Their objective was to assist in visualizing the monthly budget using tokens. We asked participants to create as many visualization alternatives as possible through tokens in 15 minutes. 
In case participants got stuck creating visualizations, we prepared four low-level analysis tasks based on the work of Amar et al.~\cite{amar2005low}: \textit{Filter}, \textit{Compute Derived Value}, \textit{Sort}, and \textit{Correlate}. We chose these four out of the ten in Amar et al.'s work because we wanted to motivate participants to physically arrange and update their visualization, rather than only mentally thinking.

\subsubsection{Ideation}
We introduced the ideation session by presenting the concept of autocomplete as it is used in search engines through a sketch (See \autoref{Sect3-Figure2}a). When people use a search engine, they open a page, input some letters to form a word, and the search engine shows some recommended completion until the user selects the desired one and submits it. We illustrated the five main steps of autocomplete systems: \textit{Start}, \textit{input}, \textit{update}, \textit{select} and \textit{output}. We presented the illustration to participants and encouraged them to apply this metaphor to the construction process of visualizations.

We then asked participants to imagine how, in the near future, a fictional smart system with no technological limitation could help them to create visualizations following the five steps of autocomplete. We asked them to present their ideas in the form of pen and paper sketches. To guide their ideation process, we  1) provided paper with a template of a scene in which someone is placing the first token on a canvas (see \autoref{Sect3-Figure2}b), and 2) displayed a list of questions on a screen while they were sketching:

\begin{enumerate}[noitemsep,topsep=0pt,parsep=0pt,partopsep=0pt]
    \item What would the empty input canvas field look like?
    \item When you put the first token, how would the system show the suggestions for alternative visualizations?
    \item When you add a new token, how would the system update its suggestions?
    \item How would you select the suggestion(s) that you want?
    \item How would the system show the final visualization after the selection is done?
\end{enumerate}

Participants had 15 minutes to sketch their answers on the sketch template(\autoref{Sect3-Figure2}c).

\subsubsection{Discussion}
Participants were asked to present and discuss their sketches with each other. They had 2-3 minutes to present their sketches to the group, explaining how a user can interact with their visualization autocomplete authoring system. After the presentation, participants were encouraged to discuss the benefits as well as the limitations of each idea. This session was video-recorded. 

\subsection{Data Analysis}
We collected 93 pages of sketches which formed 17 design ideas. 
We excluded 3 of these 17 design ideas that were autocomplete for text instead of visualization. 
To analyze the sketches, we performed a thematic analysis~\cite{braun2012thematic}. 
The authors conducted two rounds of synchronous collaborative coding sessions on the sketches with different coders. Coders were allowed to make changes to their codes between sessions to improve the identification of diverse phenomena. As each round involved a different set of coders, a third session was conducted with all coders to consolidate the codes. During each session, the coders engaged in discussions until arriving at an internal consensus. In the third round, all codes were thoroughly discussed until a consensus was reached among all coders on the definition and instances of codes. The codebook and detailed coding results are available in our online repository.

%% file: Results.tex
\section{Results}
\label{Sect4}

All participants used solid colorful squares to represent the already placed tokens. Correspondingly, participants used unfilled squares with dashed strokes to indicate the tokens yet to be placed. We describe a `step' as the operation of placing a single token on the canvas, which defines a visual mapping for that token. In general, constructing a visualization involves more than one step.  Participants produced 14 design ideas (I1-I14). From these, we identified three autocomplete strategies (\textsc{Next-Step} (2/14), \textsc{Ghost} (5/14), \textsc{Gallery} (3/14), and the remainder did not specify an autocomplete strategy). Sample sketches can be found in \autoref{Sect4-Figure1}.

\subsection{\textsc{Next-Step}}
\textsc{Next-Step} (sketches with cyan labels in \autoref{Sect4-Figure1}) bases its suggestions on proximity information of existing tokens, primarily the last placed token. The basic approach of \textsc{Next-Step} is to highlight available positions around the previous visual mapping operation, such as front, back, left, and right. I5 illustrates this approach, where the system suggests all four positions (highlight in yellow) around the previously placed red token for placing the new token. Furthermore, \textsc{Next-Step} can be more intelligent than simply showing positions in four directions. For instance, I3 ideates a similar idea with more possibilities. There are two more suggested positions at the right and the bottom that align with the previous token vertically and horizontally but are not directly connected to it. Suggestions are accepted by placing the next token, and the \textsc{Next-Step} strategy will update suggestions accordingly for the new next step. This process runs iteratively until a completed visualization forms. 

\subsection{\textsc{Ghost}}
The \textsc{Ghost} (sketches with magenta labels in \autoref{Sect4-Figure1}) strategy provides integrated suggestions that autocomplete multiple steps simultaneously. These suggestions are ghosts of a partial or completed visualization. 
I13 illustrates how the strategy would provide multiple options for partial visualizations based on the token placed. In this idea, after the user places the yellow token, the system generates three options that include a variable number of steps (The participant noted them as option 1, 2, and 3).  There is a vertical ghost bar (4 tokens yet to place) and a horizontal ghost bar (5 tokens yet to place).  Option 3 is a zigzag visualization that needs 4 more tokens. Users can then place a second token to accept one of these alternatives quickly, and the system will render the tokens automatically. 

Moreover, these options can expand to different dimensions. I10 provides ghost autocomplete options for a 2D bar chart and a 3D bar chart. Finally, once enough tokens have been placed, \textsc{Ghost} can provide suggestions for completed visualizations, as depicted in I1 and I9 (I9-1 in \autoref{Sect4-Figure1} ). The \textsc{Ghost} helps to build a 3D bar chart in I1 and autocompletes a 2D bar chart in I9. I12 shows the same idea with a more powerful system that paints the full picture with only one token placed.

\subsection{\textsc{Gallery}}

The \textsc{Gallery} (sketches with yellow labels in \autoref{Sect4-Figure1}) strategy offers integrated suggestions that autocomplete all the steps of building a visualization. The suggestions are from the perspective of the whole visualization rather than one single construction operation. \textsc{Gallery} generally provides suggestions through small multiple displays. As a top-down strategy, \textsc{Gallery} might require users to define high-level abstractions, such as data mapping, before generating multiple visual mapping suggestions. I2 and I7demonstrate this approach in their sketches. I2 requires users to map data on the left screen, and the system autocompletes visual mapping suggestions on the right screen. I7 presents visual mapping options in AR, using a virtual float screen that appears after users complete data mapping on the table. In both cases, the system automatically renders a visualization once users accept a suggestion.

\textsc{Gallery} can also be less arbitrary. It can provide suggestions without defining abstractions and allow users to finish the construction manually. For instance, in another sketch of I9 (I9-2 in \autoref{Sect4-Figure1}), the system provides a gallery of ``chart style'' options at the outset, which helps to narrow down the range of possible visualization types without automatically rendering any visualizations. Users must then construct the visualization manually, perhaps with the help of \textsc{Ghost}. 

%% file: Discussion.tex
\section{Discussion}
\label{Sect5}


\subsection{Exploring Visual Mapping Decision Tree}

\begin{figure}[h!]
\centering
\includegraphics[width = 0.9\linewidth]{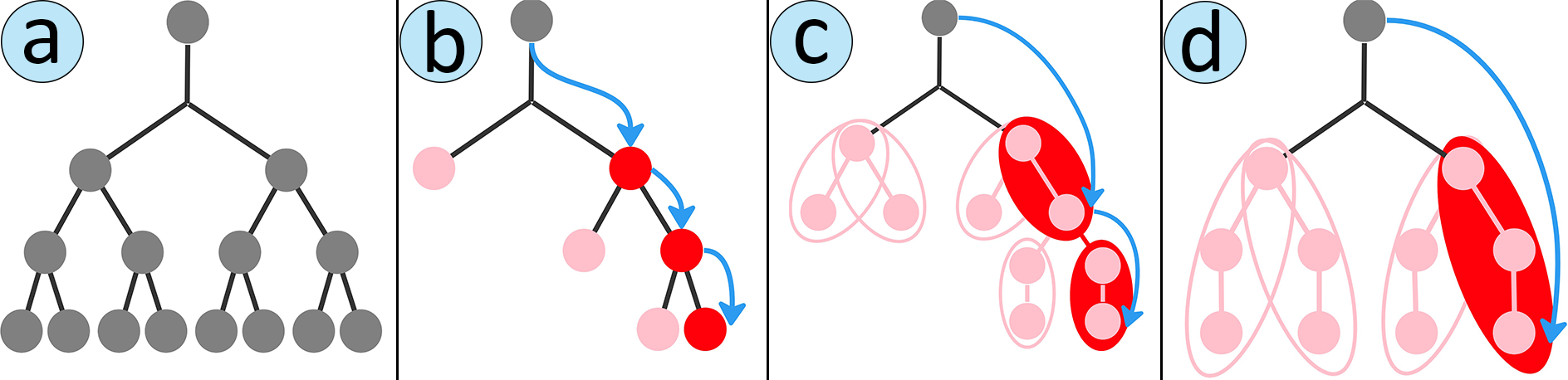}
\vspace{-0.2cm}
\caption{The different ways to explore the visual mapping decision tree. The original decision tree (a) and exploration through \textsc{Next-Step} (b), \textsc{Ghost}(c), and \textsc{Gallery} (d). } 
\label{Sect5-Figure1}
\end{figure}

The creation of visualizations through manual assembly requires defining visual mappings step by step. This process can be illustrated through a decision tree (\autoref{Sect5-Figure1}a). Each node represents a visual mapping or a rendering operation. 
A path from the root to a leaf portrays a mapping schema that defines a visualization. Each autocomplete strategy helps people to explore the decision tree in a different way.
\vspace{-0.2cm}

\vspace{-0.2cm}

Focusing on sequential steps, one at a time, \textsc{Next-Step} provides a preview for child nodes (nodes in pink, \autoref{Sect5-Figure1}b) of the next level in the decision tree. 
The preview demonstrates the possible visual mappings of the token yet to be placed. With the help of the preview, people are able to make an informed decision by accepting one node (nodes in red) in the preview at every step. By showing the choices step by step, \textsc{Next-Step} helps build a visualization from scratch, especially for non-experts. However, it still requires people to manipulate every token manually, making the authoring process time-consuming. 

The \textsc{Ghost} strategy provides suggestions for paths (marked with ellipses in \autoref{Sect5-Figure1}c) rather than individual nodes. 
By presenting suggestions integrating multiple nodes, \textsc{Ghost} allows users to define the visual mapping of multiple tokens simultaneously. As a result, \textsc{Ghost} can speed up the construction process when a desired suggestion is presented. 
However, identifying which groups of steps of the mapping or the rendering process a \textsc{Ghost} should automatize remains an open question.

Finally, \textsc{Gallery} is the ultimate version of \textsc{Ghost},  as it provides suggestions for a selection of complete paths from the root to the leaf nodes (\autoref{Sect5-Figure1}d). With this high-level of automation, \textsc{Gallery} autocompletes all visual mapping steps required for a visualization, creating a fast and hands-off experience. However, while \textsc{Gallery} is efficient, it also limits the design alternatives that manual assembly can support, and may compromise the expressive nature of the process. 

\subsection{Automation and Expressivity}
The three autocomplete strategies differ in their degree of automation. \textsc{Next-Step} requires manual input for each token. In contrast,  \textsc{Gallery} is highly automated and can exclude human participation in the construction process. \textsc{Ghost} falls in between, incorporating 
a variety of techniques composed of manual assembly and automation. In exploring design options, \textsc{Next-Step} provides the most expressivity, while \textsc{Ghost}, depending on its design, might reduce expressivity if the suggestions are inappropriate. And \textsc{Gallery} limits the exploration of choices. This trade-off suggests that increasing automation can decrease expressivity. Achieving the right balance between automation and expressivity is a complex task that requires careful consideration.

One approach is exploiting the potential of \textsc{Ghost}. \textsc{Ghost} is in between \textsc{Next-Step} and \textsc{Gallery}. This `middle' position makes it the option most likely to facilitate expressive authoring and automated construction. It is also the autocomplete strategy that appeared most frequently in our study (5/14). A perfect \textsc{Ghost} would provide recommendations that automatize the building process and allow people to make all the decisions on the visual mapping. However, identifying what part is building and what part is visual mapping is an open research and design challenge. 

Another approach is integrating multiple strategies into a single authoring tool that allows people to switch between them during different phases of visualization creation. 
For instance, when people want to play with data and explore all possibilities, \textsc{Next-Step} can serve as a `non-invasive' assistant. If people want to reduce the number of manual actions while keeping all or most of the visual mapping choices,  \textsc{Ghost} might be more appropriate. If people have a clear idea of a standard visualization, \textsc{Gallery} might be a valuable tool to expedite the visual mapping and building process. Combining these strategies can provide users with flexibility that balances both automation and expressivity.


\subsection{Limitations}

Each visualization authoring tool has benefits and constraints influencing the creation process.
Sketching on paper does not provide easy undo but allows expressivity and is accessible to the non-expert; coding provides undo and is expressive but requires expertise, while chart editors like Excel allow undoing and are less expressive and require maybe less expertise than coding and more than sketching. 
For simplicity, this study is limited to autocomplete in the context of a token assembly, which allows undoing and does not require extensive expertise. However, the conception of autocomplete of visualization or physicalizations~\cite{jansen2013interaction} at the visual mark level is way broader than using tokens or blocks. It could apply to any visualization authoring tool and paradigm. The results of our study are limited to authoring with a specific type of block. Future work could explore what strategy people will suggest with other virtual or physical authoring tools, and processes such as sketching, coding, swarm-bots, or digital fabrication~\cite{JansenOpportunitiesandChallenges2015}, visualization by demonstration~\cite{saket2016visualization}. 

%% file: Conclusion.tex
\section{Conclusion}
\label{Sect6}
The concept of visualization autocomplete was defined and introduced in this paper. In a study, we delved into how visualization autocomplete might be by using tangible tokens to construct visualizations. We identify three distinct autocomplete strategies that could aid in automating and simplifying the visual mapping and rendering process during manual visualization authoring. This study provides valuable insights for future visualization authoring tools incorporating autocomplete mechanisms, paving the way for more expressive and automated visualization authoring experiences.


